\documentclass[acmsmall]{acmart}
\pdfoutput=1
\renewcommand\footnotetextcopyrightpermission[1]{}
\usepackage{color}
\usepackage[ruled]{algorithm2e}
\usepackage{tikz}
\usepackage{pgfplots}
\usepackage{braket}
\usepackage{dsfont}
\usepackage{mathtools}
\usepackage[capitalise]{cleveref}
\usepackage{xcolor}

\definecolor{background}{HTML}{FFFFFF}
\definecolor{dataTwo}{HTML}{FF8066}
\definecolor{dataOne}{HTML}{4B4453}
\definecolor{lineColor}{HTML}{845EC2}
\definecolor{postState}{HTML}{C34A36}

\usetikzlibrary{patterns}

\setcopyright{acmcopyright}
\copyrightyear{}
\acmYear{xxxx}
\acmDOI{}

\acmJournal{TQC}
\acmVolume{xx}
\acmNumber{xx}
\acmArticle{xx}
\acmMonth{0}

\newcommand\Cmplx{\mathds{C}}

\newcommand\filltoend{\leavevmode{\unskip
  \leaders\hrule height.5ex depth\dimexpr-.5ex+0.4pt\hfill\hbox{}%
  \parfillskip=0pt\endgraf}}

\def\boxwidth{0.5}

\pgfplotsset{
    box plot/.style={
        /pgfplots/.cd,
        only marks,
        mark=-,
        mark size=\boxwidth em,
        /pgfplots/error bars/.cd,
        y dir=plus,
        y explicit,
    },
    box plot box/.style={
        line width = 0.5pt,
        /pgfplots/error bars/draw error bar/.code 2 args={%
            \draw  ##1 -- ++(\boxwidth em,0pt) |- ##2 -- ++(-\boxwidth em,0pt) |- ##1 -- cycle;
        },
        /pgfplots/table/.cd,
        y index=2,
        y error expr={\thisrowno{3}-\thisrowno{2}},
        /pgfplots/box plot
    },
    box plot top whisker/.style={
    		line width = 0.5pt,	
        /pgfplots/error bars/draw error bar/.code 2 args={%
            \pgfkeysgetvalue{/pgfplots/error bars/error mark}%
            {\pgfplotserrorbarsmark}%
            \pgfkeysgetvalue{/pgfplots/error bars/error mark options}%
            {\pgfplotserrorbarsmarkopts}%
            \path ##1 -- ##2;
        },
        /pgfplots/table/.cd,
        y index=4,
        y error expr={\thisrowno{2}-\thisrowno{4}},
        /pgfplots/box plot
    },
    box plot bottom whisker/.style={
   		 line width = 0.5pt,
        /pgfplots/error bars/draw error bar/.code 2 args={%
            \pgfkeysgetvalue{/pgfplots/error bars/error mark}%
            {\pgfplotserrorbarsmark}%
            \pgfkeysgetvalue{/pgfplots/error bars/error mark options}%
            {\pgfplotserrorbarsmarkopts}%
            \path ##1 -- ##2;
        },
        /pgfplots/table/.cd,
        y index=5,
        y error expr={\thisrowno{3}-\thisrowno{5}},
        /pgfplots/box plot
    },
    box plot median/.style={
    		mark=*,
    		mark size = \boxwidth em,
        /pgfplots/box plot
    }
}
\begin{document}

\title{Leveraging state sparsity for more efficient quantum simulations}

\author{Samuel Jaques}
\affiliation{%
	\institution{Microsoft Quantum}
	\city{Zurich}
	\country{Switzerland}
}

\author{Thomas H{\"a}ner}
\affiliation{%
  \institution{Microsoft Quantum}
  \city{Zurich}
  \country{Switzerland}
}

\begin{abstract}
	High-performance techniques to simulate quantum programs on classical hardware rely on exponentially large vectors to represent quantum states. When simulating quantum algorithms, the quantum states that occur are often sparse due to special structure in the algorithm or even in the underlying problem. We thus introduce a new simulation method that exploits this sparsity to reduce memory usage and simulation runtime. Moreover, our prototype implementation includes optimizations such as gate (re)scheduling, which amortizes data structure accesses and reduces memory usage.
	To benchmark our implementation, we run quantum algorithms for factoring, computing integer and elliptic curve discrete logarithms, and for chemistry. Our simulator successfully runs a factoring instance of a 20-bit number using 102 qubits, and elliptic curve discrete logarithm over a 10-bit curve with 110 qubits. While previous work needed a supercomputer to simulate such instances of factoring, our approach succeeds in less than 4 minutes using a single core and less than 100 MB of memory. To the best of our knowledge, we are the first to fully simulate a quantum algorithm to compute elliptic curve discrete logarithms.
\end{abstract}


\maketitle
\pagestyle{plain}
\thispagestyle{empty}

\section{Introduction}

Quantum computers promise to solve certain computational tasks exponentially faster than their classical counterparts. Speedups are expected for applications from various domains, including cryptography~\cite{shor1994algorithms} and computational chemistry~\cite{mcardle2020quantum}.

To quantify the advantage of a future quantum computer for practical applications, a host of software packages has emerged over the recent years. These tools allow domain experts to implement quantum algorithms for the purposes of verification and resource estimation~\cite{svore2018q,bichsel2020silq,steiger2018projectq,aleksandrowicz2019qiskit,green2013quipper,javadiabhari2015scaffcc}. Due to the exponential overhead in resources needed to simulate a quantum algorithm on classical hardware, verification is usually performed on individual subroutines and for small inputs. However, as quantum algorithms are being optimized, existing simulation approaches are no longer powerful enough for simulating even the smallest problem instances.

While some (non-universal) quantum circuits allow for efficient classical simulations (e.g., permutation-only circuits), state-of-the-art implementations of quantum algorithms employ certain optimizations that prohibit the use of such approaches for testing and debugging. For example, while a so-called Toffoli simulator can simulate a straightforward implementation of the modular exponentiation in Shor's algorithm for factoring, this is no longer the case when the \textit{coset}-representation optimization from the work by~\citet{zalka2006shor} is used.

Moreover, certain optimizations reduce the required quantum resources but, at the same time, reduce the algorithm's success probability, e.g., to $\mathcal O(1/N)$~\cite{gidney2019factor,roetteler2017quantum}, where $N$ grows with the problem size. For testing and debugging such implementations, simulations at nontrivial sizes are necessary.

\textbf{Related Work.} To address these issues, a variety of simulation approaches have been proposed. Full state vector simulation is advantageous when the state is dense, as is the case for random circuits. Because this approach stores all $2^n$ entries for $n$ qubits, quantum algorithms that require more than 40 qubits are challenging to simulate~\cite{haner20175}. For low-depth circuits, approaches based on tensor network contraction and a combination of (approximate) Feynman path and Schr\"odinger simulation have been shown to work best~\cite{markov2018quantum,boixo2017simulation,villalonga2020establishing,huang2020classical}.

For quantum circuits dominated by so-called Clifford gates, the method \citet{bravyi2016improved} introduced can be used. Yet, algorithms for chemistry and cryptography mostly consist of non-Clifford gates, making this approach ill-suited for testing such algorithms.
Decision-diagram based simulation~\cite{grurl2020arrays,zulehner2018advanced} performs well, e.g., for QFT and Grover search. Nevertheless, simulating full quantum algorithms with more than 40 qubits seems to be out of reach also for this method~\cite{grurl2020arrays}.

\citet{dang2019optimising} used matrix product states and an optimized entanglement mapping in order to simulate Shor's algorithm for factoring a 20-bit number in approximately 7 hours on a supercomputer. In particular, they employ emulation~\cite{haner2016high}, meaning that they directly implement modular multiplication in the simulator, instead of executing an actual implementation of modular multiplication in terms of quantum gates. However, gate-level simulation is necessary to test and debug optimized implementations (e.g., of modular multiplication).

\textbf{Contributions.} We introduce a new approach for simulating quantum algorithms that leverages sparsity in the quantum state (in a fixed basis). We present several optimizations that increase sparsity and reduce the number of write-operations to the sparse datastructure. We then benchmark our implementation by running quantum algorithms for factoring, computing elliptic curve discrete logarithms, and for quantum chemistry (both Trotter-based and qubitization-based implementations). To this end, we expose our simulator through the Q\# runtime, allowing us to rely on existing implementations of quantum algorithms in Q\#.

\begin{table*}[t]
	\caption{The largest instances of algorithms for factoring and for computing integer and elliptic curve discrete logarithm that we simulated, compared to previous state of the art. All simulations were carried out on a single core and with less than 100 MB of memory.}
	\label{tab:records}
	{\footnotesize
	\begin{tabular}{p{3cm}cccp{5cm}}
		\toprule
		Problem & Instance & Qubits & Median Runtime & Previous record\\
		\midrule
		Factoring & $N=961307$ & 102 & 219 s & \citet{dang2019optimising}: >7 hours on 216 nodes using 14 TB of RAM with \textit{emulation} \\
		EC DLog & $p=665$ & 110 & 69s& None\\
		Integer DLog &$p=32797$ & 82& 74s&None \\
		Energy estimation with qubitization  & $H_2$ with 8so &27&968s& The same simulation as~\cite{low2019qsharp} did not finish in 6 days\\
		\bottomrule
	\end{tabular}}
\end{table*}

More specifically, our contributions are as follows:
\begin{itemize}
	\item We introduce a new approach to quantum algorithm simulation that leverages sparsity in the state vector for improved performance and memory usage.
	\item We implement our simulator in C++ and expose it through the Q\# runtime.
	\item We optimize the performance of our simulator using local quantum circuit rewrites, C++ code optimizations, and multithreading.
	\item We show that our implementation is capable of simulating larger quantum algorithms than existing methods. In particular, we successfully simulate quantum algorithms for factoring, for elliptic curve discrete logarithms and for quantum chemistry that use up to 105, 110, and 27 qubits, respectively; see \cref{tab:records} for a summary of the largest simulations.
	\item To the best of our knowledge, we are the first to report a successful simulation of the quantum algorithm for computing elliptic curve discrete logarithms, see \cref{tab:records}. Even the smallest instance, a 3-bit curve, requires 40 qubits. 
	\item We present the first simulation of all of the circuit-level quantum optimizations for integer factoring  by~\citet{gidney2019factor}.
\end{itemize}

\section{Quantum Computing}
This section introduces the basic of quantum computing and classical simulation of quantum computers. For a more in-depth treatment, we refer the reader to the textbook by \citet{nielsen2002quantum}.

\subsection{Quantum States, Bits, and Gates}

Quantum computers consist of multiple quantum bits (qubits). Similar to its classical counterpart, a qubit can be in a classical state (0 or 1).

\textbf{Notation.}
We denote the 0-state by $\ket0$ (``ket zero'') and the 1-state by $\ket1$ (``ket one'').
The $\ket\cdot$-notation is part of the so-called ``bra-ket'' notation: In our setting, $\ket v$ is a column vector, whereas $\bra v$ (``bra'') denotes a row vector -- the Hermitian conjugate of $\ket v$. With this notation, the inner product between $\ket v$ and $\ket w$ can be written as $\braket{v|w}$ and their outer product is $\ket v\bra w$.

\textbf{Qubits.} The state of a single qubit $\ket\psi$ can be written as
\[
	\ket\psi =\alpha_0\ket0 + \alpha_1\ket 1,
\]
where $\alpha_0,\alpha_1\in\Cmplx$ and $|\alpha_0|^2+|\alpha_1|^2=1$. The latter is the normalization condition, which ensures that $p_i=|\alpha_i|^2$ can be viewed as a probability. Namely, $p_i$ is the probability of a measurement resulting in the outcome $i\in\{0,1\}$. By identifying $\ket0$ and $\ket 1$ with the canonical basis vectors of $\mathds C^2$, $\ket\psi$ can be written as a column vector in $\mathds C^2$,
\[
	\ket\psi = \alpha_0\begin{pmatrix}1\\0\end{pmatrix} + \alpha_1\begin{pmatrix}0\\1\end{pmatrix} = \begin{pmatrix}\alpha_0\\\alpha_1\end{pmatrix}.
\]

Analogously, the state of an $n$-qubit system is a complex superposition over all $n$-bit strings:
\[
	\sum_{i=0}^{2^n-1} \alpha_i\ket i,
\]
where $\ket i := \ket{i_{n-1}}\otimes \cdots\otimes\ket{i_0}$ and $i_k$ denotes the $k$-th bit of the integer $i$ and $\otimes$ is the Kronecker product. As before, the corresponding column vector in $\mathds C^{2^n}$ has entries $\alpha_i$.

\textbf{Quantum Gates.} Quantum computation is performed by applying operations -- so-called quantum gates -- to qubits. We distinguish gates from measurements. A measurement of all $n$ qubits results in the outcome $i=i_{n-1},...,i_0$ with probability $p_i = |\alpha_i|^2$. A quantum gate that acts on $n$ qubits can be represented as a $2^n\times 2^n$-dimensional complex matrix, and gate application is modeled as matrix-vector multiplication. In contrast to measurements, which project the state onto the observed outcome (i.e., $\ket i$ will be the post-measurement state when $i$ was observed), gates must be unitary. That is, the matrix $U$ describing a gate must satisfy $U^\dagger U=UU^\dagger=\mathds 1$, where $U^\dagger$ is the Hermitian conjugate of $U$.

Some common single-qubit gates and their respective matrices are the Hadamard gate $H=\sqrt{0.5}\begin{psmallmatrix}1&1\\1&-1\end{psmallmatrix}$, the $S$ gate $S=\begin{psmallmatrix}1&0\\0&i\end{psmallmatrix}$, the Pauli gates $X=\begin{psmallmatrix}0&1\\1&0\end{psmallmatrix}, Y=\begin{psmallmatrix}0&-i\\i&0\end{psmallmatrix},Z=\begin{psmallmatrix}1&0\\0&-1\end{psmallmatrix}$, and Pauli rotation gates $R_P(\theta)=e^{-0.5i\theta P}$, where $P\in\{X,Y,Z\}$.

An example of a two-qubit gate is a controlled single-qubit gate, which only applies the single-qubit gate if a control qubit is in $\ket 1$. Since $\ket i\bra i$ is the projector onto $\ket i$, we can write a controlled single-qubit gate $U$ as
\[
	CU = \ket0\bra0\otimes\mathds 1 + \ket1\bra1\otimes U.
\]
We note that this definition may be iterated to construct $n$-ary controlled gates, which we denote by C$^n U$.

There are several common quantum operations which just permute computational basis states, meaning they are equivalent to reversible classical operations. This includes the Pauli X or NOT gate, swap, CNOT, and C$^n$NOT, which denotes an $n$-ary controlled NOT. 

\subsection{Classical Simulation}
A natural method to simulate a quantum computer is the Schr{\"o}dinger full-state simulation, where an array of $2^n$ complex numbers represents the quantum state. The coefficient of state $\ket{i}$ is stored at location $i$ in the array. This method is widely used in the literature~\cite{smelyanskiy2016qhipster,jones2019quest,haner20175}.

One advantage of this approach is its generality. Any quantum operation, including noisy operations, can be simulated using this method. Unfortunately, the space requirement is exponential in the number of qubits and going significantly beyond 45 qubits is impractical~\cite{haner20175}.

While actual quantum computers will also face strict qubit limits, we expect more flexible space-time trade-offs. One reason is that longer, more complex algorithms will require more error correction which, in turn, increases the overhead in terms of the number of physical qubits. In addition, the trade-offs may be very one-sided. For example, new algorithms for quantum chemistry can reduce gate complexity by a factor of 1000 with only a factor of 20 more qubits~\cite{berry2019qubitization}, and for factoring integers, 50\% more qubits can reduce the gate count by 2000 times \cite{gidney2019factor}.
These algorithmic improvements thus significantly reduce the runtime of a quantum algorithm. However, the increase in qubit numbers makes classical testing and debugging a challenging task. Approximate arithmetic adds $k$ qubits and fails with probability $\Omega(2^{-k})$, requiring many qubits to simulate modestly accurate results. The lowest number of qubits for computing elliptic curve discrete logarithms~\cite{haner2020improved} is estimated at $8n+O(\lg n)$. We find the smallest possible $n=3$ uses 40 qubits. At this size, the point addition circuits will fail for 3/6 points on the curve.

\section{Sparse States for Simulation}

Many quantum algorithms with space-time trade-offs will result in states of the form 
\begin{equation}
\sum_{i\in\{0,1\}^n}\alpha_i\ket{i, f(i)}
\end{equation}
for some function $f:\{0,1\}^n\rightarrow\{0,1\}^m$. This state belongs to a vector space of dimension $2^{n+m}$, but it is a sparse vector, at least in the computational basis. This means full-state simulators will waste both memory and time on an exponentially large number of zero-amplitudes.

Our new approach makes use of this sparsity by only storing nonzero entries, thus reducing both memory usage and simulation runtime. Specifically, our simulator stores a quantum state $\sum_{b\in\{0,1\}^n}\alpha_b\ket{b}$ as unordered key-value pairs $(b,\alpha_b)$ for all $\alpha_b\neq 0$, for which we use the efficient hash map by~\citet{skarupke2018hash}. We refer to $b$ as the label and $\alpha_b$ as the amplitude. 

Compared to a full-state simulator, we use slightly more memory to represent each amplitude, as we store not only the label, but 1 byte for metadata in the hash map. However, we save memory for every zero-amplitude in the corresponding full state representation.

\subsection{Operations}
We implement the intrinsic operations in the Q\# language. These include Pauli gates, phase rotations, Pauli rotations, measurements, and Pauli exponentials, as well as multiply-controlled versions of all the above (besides measurements). A useful feature of all of these operations is that, when expressed as matrices, at most 2 entries in each row and column are non-zero. We conceptually divide the operations into those with at most 1 non-zero entry, and others. 

For the second category, performing an operation requires some computation on pairs of basis states $\ket{b}$ and $\ket{b'}$. We can represent the action of the operator on the space that $\ket{b}$ and $\ket{b'}$ span as 
\begin{equation}
\begin{pmatrix}\alpha'_{b} \\ \alpha'_{b'}\end{pmatrix} =
\begin{pmatrix} a_{11}(b) & a_{12}(b)\\ a_{21}(b) & a_{22}(b)\end{pmatrix} \begin{pmatrix} \alpha_b \\ \alpha_{b'}\end{pmatrix}
\end{equation}
where $a_{ij}$ are easily computable functions of $b$. For all the quantum gates we implement, $b'$ is easy to compute from $b$ and vice versa.

To compute an operation, we iterate through all label-amplitude pairs in the hash map. For each one, we compute new label-amplitude pair(s) and insert them into a new hash map. This requires at most one look-up in the original hash map because each label has at most one partner. Since each pair of partner labels differs in at least one bit, if both are present in the table, we only insert when the iteration reaches the label with $1$ at this bit location. 

While the use of a new hash map doubles the memory requirements, an online algorithm to modify an existing hash map in-place would be challenging. A major issue is that most operations would drastically change the labels, which could force re-sizing or re-hashing. Preserving the ordering of an iteration in an unordered map during a re-hashing is a difficult problem that we did not attempt to solve. However, a benefit of our simple approach is that we can queue operations with exactly one non-zero entry.

\subsection{Queueing phases and permutations}
Because the arithmetic intensity of each gate is low, a large portion of the computation is spent on look-ups, insertions, and initialization of the new hash map. To minimize these overheads, we queue gates where possible. 

As matrices, Pauli gates and phase rotations with any number of controls only have a single non-zero entry per row. We will refer to these as phase/permutation gates. Their structure allows us to define such a gate $G$ through functions $f$ and $g$ such that $G(\alpha_b\ket{b}) = f(b)\alpha_b\ket{g(b)}$. Such operations can be chained together. If $G_1$ and $G_2$ are defined by functions $(f_1,g_1)$ and $(f_2,g_2)$, we can evaluate $G_2\circ G_1(\alpha_b\ket{b}) = f_1(b)f_2(g_1(b))\ket{g_2(g_1(b))}$, with a similar pattern for any number of these gates.

To evaluate a queue of $n$ phase/permutation gates, we iterate through the hash map as before. For each pair $(b,\alpha_b)$, we compute the action of all $n$ gates before inserting the result into the new hash map. This amortizes the overheads of hash map look-ups, insertions, and initializations. 

\subsection{Gate commuting} 
For two types of quantum gates $G_1$ and $G_2$, there are often well-known and simple relationships between the serial applications $G_1\circ G_2$ and $G_2\circ G_1$, known as commutation relations. Using these relations, a circuit can be arbitrarily rewritten by altering the order of gates. 

\begin{figure}
\resizebox{0.6\textwidth}{!}{
\begin{tikzpicture}
\node at (0,1.6) (Ry) {$R_y$};
\node at (0,0.8) (Rx) {$R_x$};
\node at (0,0) (H) {$H$};

\foreach \j in {0,...,2}{
\foreach \i in {1,...,8}{
	\draw (-0.4+\i,-0.25+0.8*\j) rectangle (0.4+\i,0.25+0.8*\j) node[pos=0.5] {$0$};
}
}
\draw[fill=black] (1.6,1.6-0.25) rectangle (2.4, 1.6+0.25) node[white,pos=0.5] {${\color{postState}-}0.6$};

\node[rectangle,draw=black] at (1, 2.7) (H0) {$H$};
\node[rectangle,draw=black] at (2,2.7) (H1) {$H$};
\node[rectangle,draw=black] at (3,2.7) (X2) {$X$};
\draw[fill=black] (4,2.7) circle (0.1);
\draw[fill=black] (5,2.7) circle (0.1);
\draw 			  (6,2.7) circle (0.1);
\draw (4,2.7) -- (6.1, 2.7);
\draw (6,2.6) -- (6, 2.8);
\node at (1,3.2) {(a)};
\node at (2,3.2) {(b)};
\node at (3,3.2) {(c)};
\node at (5,3.2) {(d)};
\node at (7,3.2) {(e)};
\draw[postState,line width=0.7,->] (H0)--(1,0.3);
\draw[postState,line width=0.7,->] (H1) -- (2,1.9);
\draw[postState,line width=0.7,->] (2,1.3)--(2,0.3);

\draw[postState,line width=0.7,->] (X2) -- (3,0.3);
\draw[postState,line width=0.7,->] (3,-0.3) -- (3,-1.25) -- (4.6,-1.25);
\draw[postState,line width=0.7,->] (4,2.55) -- (4,1.2);
\draw[postState,line width=0.7,->] (5,2.55) -- (5,1.2);
\draw[postState,line width=0.7,->] (6,2.55) -- (6,1.2);
\draw[postState,line width=2] (4.6,1.15) -- (5.4,1.15);

\draw[fill=black] (1.6,-0.25) rectangle (2.4, 0.25) node[pos=0.5,white] {1};
\draw[fill=black] (2.6,-0.25) rectangle (3.4, 0.25) node[pos=0.5,white] {1};
\draw[fill=black] (3.6,-0.25) rectangle (4.4, 0.25) node[pos=0.5,white] {1};
\draw[fill=black] (4.6,-0.25) rectangle (5.4, 0.25) node[pos=0.5,white] {1};
\draw[pattern=crosshatch,pattern color=postState] (0.6,-0.25) rectangle (1.4,0.25) node[pos=0.5, black] {0};
\draw[pattern=crosshatch,pattern color=postState] (1.6,-0.25) rectangle (2.4,0.25) node[pos=0.5, white] {1};

\node[right] at (0,-1.25) (queue) {Queue};

\draw (1.5,-2.05) rectangle (7.5, -0.45);
\draw (6.8,-2) rectangle (7.4,-0.5) node[pos=0.5] {\rotatebox{90}{CNOT$_{0,3}$}};
\draw (6.1,-2) rectangle (6.7,-0.5) node[pos=0.5] {\rotatebox{90}{Z$_2$}};
\draw (5.4,-2) rectangle (6.0,-0.5) node[pos=0.5] {\rotatebox{90}{CZ$_{1,4}$}};
\draw[postState] (4.7,-2) rectangle (5.3,-0.5) node[pos=0.5,postState] {\rotatebox{90}{Z$_{2}$}};

\draw[fill=black] (6.6, 0.8-0.25) rectangle (7.4,0.8+0.25) node[pos=0.5, white] {$-\pi/4$};
\draw[fill=black] (4.6, 0.8-0.25) rectangle (5.4,0.8+0.25) node[pos=0.5, white] {$0.75$};

\end{tikzpicture}

}
\caption{Illustration of gate queues and commuting on 8 qubits. Instructions from Q\# come in at the top, each column correspons to a qubit, and the phase/permutation gate queue is at the bottom. Original state is in black. (a) Q\# sends an $H$ gate to qubit 0. Since no other gates are queued, it will set the $H$ queue on qubit 0 to 1. (b) Q\# sends an $H$ gate to qubit 1. Since $HR_y(\theta)=R_y(-\theta)H$, it will flip the sign of the angle of the $R_y$ gate. Because $H^2=I$ (the no-op gate), it will reset the $H$ queue on qubit 1 to 0 and not modify the state vector. (c) Since $XH=HZ$, it will add a $Z$ gate (in red) to the phase/permutation queue and not change the $H$ queue. (d) Q\# sends a CCNOT gate to qubits 3,4, and 5, but the simulator cannot commute this gate through the queued $R_x$ gate (nor the $H$ gates), so it first sends all the gates in the phase/permutation queue to modify the state vector. It then applies both $H$ gates on qubits 3 and 4 to the simulator, then the $R_x(0.75)$ rotation. Since all queues are clear, it then adds the original CCNOT to the (now empty) phase/permutation queue. (e) An $R_x(-\pi/4)$ gate might double the superposition size, but the simulator does not apply it during any of the previous steps.}
\label{fig:commuting}
\end{figure}
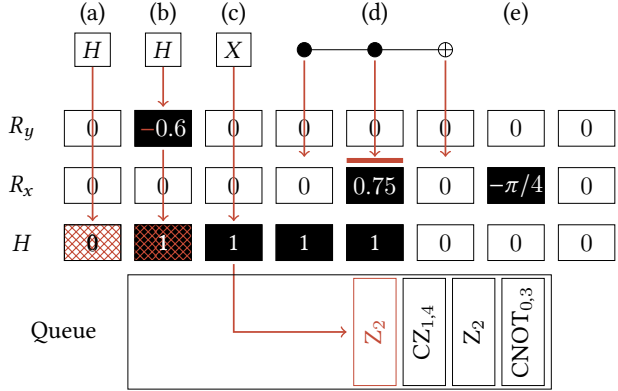

We use simple strategy with queues of four types of gates: Phase/ permutation gates, $H$, $R_x$, and $R_y$. When the Q\# runtime sends a gate $G$ to the simulator, it checks for any $R_y$, $R_x$, then $H$ gates queued on the requested qubit, and attempts to commute $G$ through each in turn. If a simple commutation relation exists, it modifies $G$ as necessary and continues. If, on the other hand, the commutation relation is difficult to compute, it executes all queued gates on those qubits, including any phase/permutation gates. \cref{fig:commuting} illustrates this process and, in particular, the interaction between gate commuting and the queue.

Besides the possibility of canceling gates and never needing to execute them, this saves memory by postponing $H$, $R_x$, and $R_y$ gates, each of which may increase the number of states in superposition.

\section{Algorithm Benchmarks}\label{sec:benchmarks}
In order to benchmark our simulator, we choose three variants of Shor's algorithm and two quantum chemistry algorithms. We run all but one type benchmark 30 times on a single core of an Intel Xeon Platinum 8168 CPU at 2.70 GHz with 32 GiB of RAM. For the quantum-optimized factoring implementation, we used 8 cores on the same machine and ran it 200 times.

Because our simulator's runtime is almost independent of the number of qubits, we did not always choose implementations with the fewest possible qubits. 

\subsection{Integer factoring}

\textbf{Background.}
Shor's algorithm for integer factoring~\cite{shor1994algorithms} has attracted a lot of attention because of its impact on cryptography. Much of the previous quantum circuit design literature focuses on the subroutine which performs modular exponentiation in superposition~\cite{beauregard2003circuit,
gidney2017factoring,gidney2018halving,
gidney2019approximate,gidney2019factor,
gidney2019windowed,haner2017factoring,
vanmeter2005fast,zalka2006shor}, since this is the most expensive part. 

The method with the fewest qubits uses Fourier-basis quantum arithmetic. However, this requires more quantum gates than other techniques and so other analyses use reversible classical addition. Recent techniques use even more qubits to drastically reduce both circuit size and depth, using quantum modifications of reversible classical techniques~\cite{gidney2019factor}. 

The semi-classical quantum Fourier transform~\cite{griffiths1996semiclassical} is standard for Shor's algorithm, and reduces the number of qubits by $2n-1$ to factor an $n$-bit integer. For our simulator, this is particularly effective, as we only end up with a superposition of all powers of a generator and one control qubit, rather than a superposition of $2^{2n+1}$ control states. 

\textbf{Setup.}
We use a semi-classical Fourier transform with a reversible classical adder (``CDKM adder'')~\cite{cuccaro2004new} and a windowed style of in-place modular multiplication~\cite{gidney2019windowed}. For comparison, we also simulate Fourier-basis quantum arithmetic (``QFT adder'')~\cite{draper2000addition}. We use an implementation of Shor's algorithm that uses $2n+1$ bits for the phase estimation.

Since the size of the superposition depends on the order of a randomly-chosen generator, we pre-selected a generator of maximal order to benchmark worst-case performance.

We also implemented all of the quantum optimizations of~\cite{gidney2019factor}: coset representations, windowed exponentation and modular multiplication, and oblivious carry runways. We used one oblivious carry runway per integer, and added the runway into the entire register before multiplication (see discussion). Since this circuit failed so frequently, we ran 200 benchmarks for each modulus, of which only 82 and 112 (respectively) finished successfully.

\begin{figure}
\centering
\begin{tikzpicture}
  \begin{axis} [axis background/.style={fill=background},enlarge x limits = 0.1, xmode=log, log ticks with fixed point, ymode=log,
  		xlabel = {Factored integer}, ylabel = {Runtime (s)}, legend pos = south east, empty legend]
    \addplot [draw=dataOne,box plot median] table {factoring.dat};
    \addplot [draw=dataOne,box plot box] table {factoring.dat};
    \addplot [draw=dataOne,box plot top whisker] table {factoring.dat};
    \addplot [draw=dataOne,box plot bottom whisker] table {factoring.dat};
    \addlegendentry[dataOne]{CDKM Adder}
    \addplot [draw=dataTwo, box plot median] table {QFTfactoring.dat};
    \addplot [draw=dataTwo,box plot box] table {QFTfactoring.dat};
    \addplot [draw=dataTwo,box plot top whisker] table {QFTfactoring.dat};
    \addplot [draw=dataTwo,box plot bottom whisker] table {QFTfactoring.dat};
    \addlegendentry[dataTwo,mark=*]{\textbf{QFT Adder}}
    \addplot [draw=lineColor, box plot median] table {QOptfactoring.dat};
    \addplot [draw=lineColor,box plot box] table {QOptfactoring.dat};
    \addplot [draw=lineColor,box plot top whisker] table {QOptfactoring.dat};
    \addplot [draw=lineColor,box plot bottom whisker] table {QOptfactoring.dat};
    \addlegendentry[lineColor,mark=*]{\textbf{Quantum Optimizations}}
   \end{axis}
\end{tikzpicture}%
\caption{Box plots from simulations of integer factoring with Shor's algorithm. Boxes have a minimum thickness for readability.}\label{fig:factoring-instances}
\end{figure}
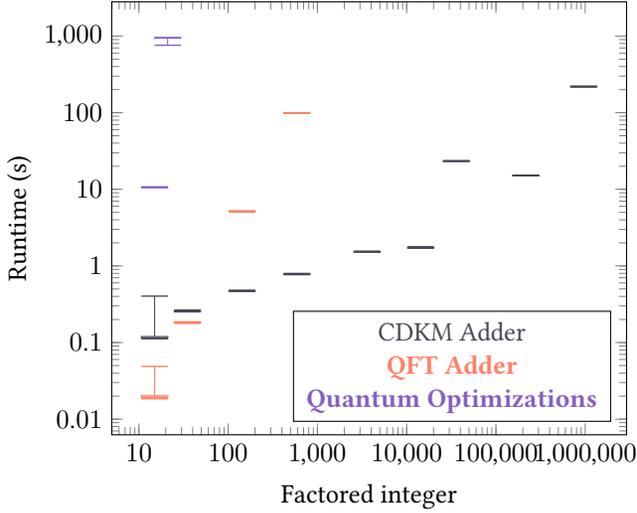

\textbf{Results.}
\cref{tab:factoring-instances}, \cref{tab:factoring-instances-quantum}, and \cref{fig:factoring-instances} show our results. Without quantum-specific optimizations, the maximum size of the superposition is exactly double the group order, as expected. The number of qubits is $5n+2$ for the CDKM adder and $2n+3$ for the QFT adder. Even in the QFT adder, only 2\% of the computational basis states are non-zero at $N=589$, though the CDKM adder uses less than $10^{-11}$\%.

Quantum-specific optimizations such as the coset representation and oblivious carry runways result in larger superpositions, as can be seen in \cref{tab:factoring-instances-quantum}. Consequently, simulations are challenging even for small instances.

\graphicspath{{./}}

\begin{table*}
\caption{Parameters of the factoring instances. $N=pq$ is the integer being factored, of the given bit-size. The group order is $(p-1)(q-1)/gcd(p-1,q-1)$. }
\label{tab:factoring-instances}
{\footnotesize
\begin{tabular}{rrr rr rr}
\toprule
	& 			&  			& 	\multicolumn{2}{c}{\filltoend\hspace{0.2em}  CDKM adder \hspace{0.2em}\filltoend}	& \multicolumn{2}{c}{\filltoend\hspace{0.2em}  QFT adder \hspace{0.2em}\filltoend}	\\
$N$ & Bit-size & Group Order & Qubits & Max. state size  & Qubits & Max. state size\\
\midrule
15 & 4 & 4 & 22	&8 & 11 & 128\\
35 & 6 & 12 & 32	& 24 & 15 & 1536\\
143 & 8 & 60  & 42	&120 & 19 & 30720\\
589 & 10 & 90  &52	&180 & 23 & 184320\\
3599 & 12 & 1740 &62	&3480 & -- & -- \\
14351 & 14 & 1008  & 72	&2016  & -- & -- \\
36089& 16 & 17850 & 82&	35700 & -- & -- \\
216067 & 18 &35856  & 92	&71712 & -- & -- \\
961307 & 20 &  479568 & 102 & 959136 & -- & -- \\
\bottomrule
\end{tabular}}
\end{table*}

\begin{table*}
\caption{Parameters of factoring instances with all quantum optimizations from~\citet{gidney2019factor}. }
\label{tab:factoring-instances-quantum}
{\footnotesize
\begin{tabular}{rrr rrr rr rr}
\toprule
	& 			&  	Group		& Coset/Runway &  	\multicolumn{2}{c}{\filltoend\hspace{0.2em}  Window Sizes \hspace{0.2em}\filltoend}	&	\multicolumn{2}{c}{\filltoend\hspace{0.2em}  Success Rate \hspace{0.2em}\filltoend}	& &	\\
$N$ & Bit-size & Order & Padding& Exp. & Mult. & Exp.& Actual  & Qubits & Max. state size\\
\midrule
15 & 4 & 4 & 4 & 3 & 4 & 0.30 & 0.41 & 40 & 557056\\
21 & 5 & 6 & 5 & 3 & 5 & 0.46 & 0.56 & 49 & 25952256\\
\bottomrule
\end{tabular}}
\end{table*}

\textbf{Discussion.}
The runtime of our simulator scales almost linearly with the group order when the CDKM adder is used, since each gate requires time linear in the size of the superposition, which is double the group order in this case.
Fourier-basis arithmetic poses a greater challenge for our simulator because it represents integers with a superposition of computational basis states. The states in this algorithm are thus much more dense.

Our simulator clearly outperforms other simulators when using the CDKM adder. Factoring 961307 took 7.2 hours on a 216 node cluster using matrix product states~\cite{dang2019optimising}, though this used a full quantum Fourier transform (QFT) and emulation for the modular arithmetic. A full QFT would be more difficult for our simulator, since it would create $\approx 2^{2n+1}\approx N^2$ states in superposition, rather than the $\approx 2N$ that we found; however, a full gate-level simulation would be more challenging for the matrix product state simulator. 
The decision-diagram simulator by~\citet{grurl2020arrays} took approximately 20 s to simulate factoring 187 with 34 qubits. Our implementation uses more qubits for a roughly equivalent problem (factoring 143), but still simulates this with a median runtime of 0.47 s. Even with Fourier-basis arithmetic, our simulator's median runtime was 5.1 s.

The quantum-optimized circuit poses a much greater challenge for our simulator in terms of both runtime and success probability. This is not surprising, given that it relies on large superpositions and, while the error rate scales exponentially with the number of padding bits, so does the superposition size and hence the memory requirements and runtime of our simulator.

Each addition adds an error of approximately $2/2^{c_{pad}}$~\cite{gidney2019approximate}, where $c_{pad}$ is the number of padding bits, and the factor of 2 accounts for both coset representation and oblivious carry runway errors. This means even a smaller factoring sample, which may require only a few dozen additions, quickly becomes lost to noise. To benchmark any examples successfully, we thus increase the window sizes for exponentiation and multiplication, which reduces the number of additions to 12 and 16 for factoring 15 and 21, respectively. 

We decode and measure all integers in between multiplications and abort if the measurements are non-zero. This saves the simulator from storing unnecessary superpositions. Each window of each multiplication involves 2 additions; however, the first of these is into an all-zero register, so it cannot cause an error in the carry runways or the coset. Thus we expect 9 and 12 additions that could actually cause errors for the two benchmarks. The error rate of each addition should be $1/2^3$ and $1/2^4$, respectively, leading to expected success rates of $(1-2^{-3})^9$ and $(1-2^{-4})^{12}$. These are both lower than what we found.

\textbf{Oblivious Carry Runways.} Modular multiplication uses the individual bits of an integer (possibly several at once) to control addition of multiples of a constant. This works because if an integer $x$ is represent by bits $(x_0,x_1,\dots, x_n)$ such that
\begin{equation}
x = x_0+2x_1+2^2x_2+\dots +2^nx_n
\end{equation}
then we can compute $bx\mod N$ as 
\begin{equation}
\left(x_0(b\mod N) + x_1(2b\mod N) + x_2(2^2b\mod N)+\dots+x_n(2^nb\mod N)\right)\mod N.
\end{equation}
This relies on the modulo operation acting as a homomorphism between addition over the integers to addition modulo $N$. With a carry runway, the integer $x$ is represented by $k$-bit words $(x_0,x_1,\dots x_n)$ and runways $(r_0,\dots, r_{n-1})$ such that
\begin{equation}
x = x_0 + 2^kr_0 + (x_1-r_0)2^k + \dots + (x_n-r_{n-1})2^{kn}.
\end{equation}
This could be used directly to control multiplication modulo $N$, though it will require more additions than necessary because of the extra runway bits. However, it's not clear how $x_i-r_{i-1}$ is represented, since this number may be negative. The approach by \citet{gidney2019approximate} is to perform this subtraction modulo $2^k$. This means we cannot perform a bit-by-bit controlled addition of multiples of a constant $b$, because if $(x_i-r_{i-1})\mod 2^k$ is represented as a bitstring $y_0+2y_1+\dots + 2^{k-1}y_{k-1}$, then it may occur that
\begin{equation}
\left(b(x_i-r_{i-1})\mod 2^k\right)\mod N \neq \left(y_0(2^0b\mod N) + \dots + y_{k-1}(2^{k-1}b\mod N)\right)\mod N
\end{equation}
because addition modulo $2^k$ is not homomorphic to addition modulo $N$.  

When $x_i \geq r_{i-1}$, there is no issue since $x_i - r_{i-1}\mod 2^k = x_i - r_{i-1}$ as integers. \citet{gidney2019factor} proposed a runway size of only around 30 qubits, and the register contains more than 1000 qubits, so the number of states with $x_i < r_{i-1}$ is negligibly small and the technique works without issue.

However, when we implemented these circuits, we had to use runways that were the same size as the register including coset padding, meaning $x_i < r_{i-1}$ for approximately half the states in superposition. To reduce the probability of failure, we had to fully propagate and decode the runways into the registers before using them as inputs into multiplications. This adds only a small overhead in depth compared to the optimized decoding by \citet{gidney2019factor}.

\subsection{Integer Discrete Logarithms}

\textbf{Background.}
Shor's algorithm also applies to finding discrete logarithms in abelian groups. We tested this for the multiplicative of integers modulo a prime number. The IETF standard~\cite{gilmor2016negotiated}, the Microsoft Schannel Provider, and ElectionGuard~\cite{benaloh2020electionguard} rely on the difficulty of discrete logarithms in this setting. 

For Shor's algorithm, the main difference compared to factoring is that we multiply by powers of two different elements of the group, and the result consists of two numbers $(j,k)$ such that $j+dk\approx 0\mod 2^m$, where $m$ is the number of bits we used in the phase estimation, and $d$ is the discrete logarithm.

\textbf{Setup.}
We use the same arithmetic circuits as our integer factoring benchmarks, with a semi-classical quantum Fourier transform for the phase estimation, and a generator of maximal order. In all benchmarks we used a discrete logarithm of 7.

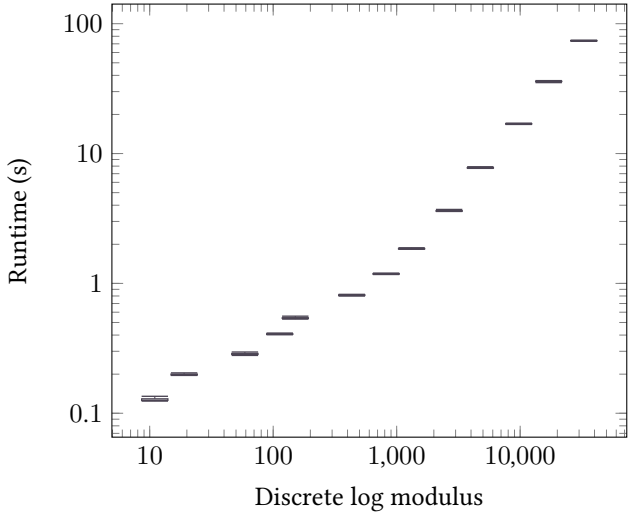
\begin{figure}
\centering
\begin{tikzpicture}
  \begin{axis} [enlarge x limits = 0.1, xmode=log, log ticks with fixed point, ymode=log,
  		xlabel = {Discrete log modulus}, ylabel = {Runtime (s)}, legend pos = south east, empty legend]
    \addplot [draw=dataOne,box plot median] table {intdlog.dat};
    \addplot [draw=dataOne,box plot box] table {intdlog.dat};
    \addplot [draw=dataOne,box plot top whisker] table {intdlog.dat};
    \addplot [draw=dataOne,box plot bottom whisker] table {intdlog.dat};
   \end{axis}
\end{tikzpicture}%
\caption{Box plots of simulations of integer discrete logarithm with Shor's algorithm. Boxes have a minimum thickness for readability.}\label{fig:int-dlog}
\end{figure}

\textbf{Results.}
\cref{tab:int-dlog-instances} and \cref{fig:int-dlog} show the runtimes of our simulator. We see a similar qubit count of $5n+2$ for $n$-bit primes.

\begin{table}
\caption{Parameters of integer discrete logarithm simulations. $p$ is the prime modulus of the given bit-size, and the group order is $p-1$. }
\label{tab:int-dlog-instances}
{\footnotesize
\begin{tabular}{rrrrr}
\toprule
$p$ & Bit-size & Group Order & Qubits & Max. state size \\
\midrule
11&4 & 10 & 22 & 40 \\ 
19&5 & 18 & 27 & 72 \\ 
59&6 & 58 & 32 & 232 \\ 
113&7 & 112 & 37 & 448 \\ 
151&8 & 150 & 42 & 600 \\ 
433&9 & 432 & 47 & 1728 \\ 
821&10 & 820 & 52 & 3280 \\ 
1321&11 & 1320 & 57 & 5280 \\ 
2659&12 & 2658 & 62 & 10632 \\ 
4759&13 & 4758 & 67 & 19032 \\ 
9743&14 & 9742 & 72 & 38968 \\ 
17033&15 & 17032 & 77 & 68128 \\ 
32797&16 & 32796 & 82 & 131184 \\ 
\bottomrule
\end{tabular}}
\end{table}

\textbf{Discussion.}
The performance is very similar to integer factoring, since both are simply controlled modular multiplication, from the perspective of the simulator. 

It may seem surprising that the state size is 4 times the group order, whereas it was $2\times$ for factoring. The reason is that the semi-classical QFT is performed from the most-significant bit to the least significant, and the group order of RSA-style moduli is always even. Thus, until the final modular multiplication, the superposition only contains powers of the \emph{square} of the generator, not the generator itself. In contrast, integer discrete logarithm performs a full semi-classical QFT on the generator before moving to the target integer. After the full semi-classical QFT, the full group is present in superposition. 

The particular implementation of modular addition we used performs a measurement-based uncomputation of a comparison, which temporarily doubles the state size. Thus, since integer discrete logarithms continue to perform modular multiplication after creating a full group superposition, but factoring does not, the maximum superposition size in factoring is only twice the group order.

\subsection{Elliptic curve discrete logarithm}

\textbf{Background.}
Another major application of Shor's algorithm to cryptography is solving the discrete logarithm problem for elliptic curve groups. The structure of Shor's algorithm is the same as integer discrete logarithms, but the arithmetic of the underlying group is much more involved. In particular, the arithmetic requires many more auxiliary registers, meaning that a full-state simulation would be limited to curves of only 2-3 qubits.

\textbf{Setup.}
We used a mix of ``low-width'' and ``low-T'' elliptic curve operations from the Q\# implementation by~\citet{haner2020improved}, with a set of prime-order curves defined over primes of 5 to 10 bits. We also used the signed windowed quantum Fourier transform, with a window size equal to $\lfloor\lg n\rfloor$ for $n$-bit primes.

The quantum circuit for elliptic curve point addition adds a classical point $P$ to a quantum point $Q$, but fails if $P=Q$, $P=-Q$, or $P+Q=-P$. The behaviour in these cases results in states representing ``points'' which are not on the elliptic curve.  Once a point is not on the curve, the behaviour of the addition formulas is not defined, and they may produce more pairs of points that are not on the curve. 

For an elliptic curve over a prime $p$, the order is $p\pm O(\sqrt{p})$~\cite{hasse1936zur}, but there are $p^2$ possible pairs of points $(x,y)$ modulo $p$. Thus, fully simulating this imperfect point addition creates far more states in superposition than a perfect point addition. We measured an auxiliary register in the addition circuit which is returned to 0 for well-defined point addition. This either removes some of the undesired states, or causes a complete failure of the addition. In the second case we abort the benchmark and do not include it in the results.

\begin{figure}
\centering
\begin{tikzpicture}
  \begin{axis} [enlarge x limits = 0.1, xmode=log, log ticks with fixed point, ymode=log,
  		xlabel = {Discrete log modulus}, ylabel = {Runtime (s)}, legend pos = south east, empty legend]
    \addplot [draw=dataOne,box plot median] table {eccdlog.dat};
    \addplot [draw=dataOne,box plot box] table {eccdlog.dat};
    \addplot [draw=dataOne,box plot top whisker] table {eccdlog.dat};
    \addplot [draw=dataOne,box plot bottom whisker] table {eccdlog.dat};
    \addlegendentry[dataOne]{Semi-classical QFT}
    \addplot [draw=dataTwo, box plot median] table {eccdlogwindowed.dat};
    \addplot [draw=dataTwo,box plot box] table {eccdlogwindowed.dat};
    \addplot [draw=dataTwo,box plot top whisker] table {eccdlogwindowed.dat};
    \addplot [draw=dataTwo,box plot bottom whisker] table {eccdlogwindowed.dat};
    \addlegendentry[dataTwo]{\textbf{Windowed QFT}}
   \end{axis}
\end{tikzpicture}%
\caption{Box plot of simulations of elliptic curve discrete logarithm with Shor's algorithm. Boxes have a minimum thickness for readability.}\label{fig:ecc-dlog}
\end{figure}
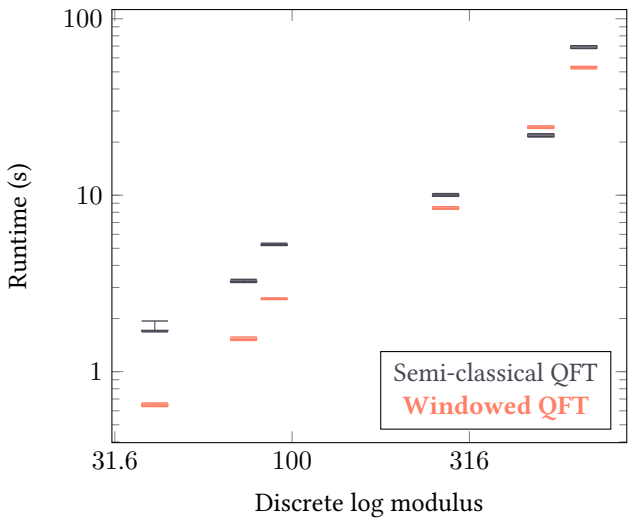

\textbf{Results.} \cref{tab:ecc-dlog-instances} and \cref{fig:ecc-dlog} show the results. We find the number of required qubits for an $n$-bit prime are $9n+\lfloor\lg n\rfloor +12$.

\begin{table*}
\caption{Parameters of the elliptic curves we tested. ``Failures'' is the number of instances where the measurement of the auxiliary register yielded non-zero and we stopped the test. The maximum state size is the maximum over all runs.}
\label{tab:ecc-dlog-instances}
{\footnotesize
\begin{tabular}{rrrr rrr rrrr}
\toprule
 	  &  Bit 	     & Group	&Exp.&  \multicolumn{3}{c}{\filltoend\hspace{0.2em}  Semi-classical QFT \hspace{0.2em}\filltoend} & \multicolumn{4}{c}{\filltoend\hspace{0.2em} Windowed QFT\hspace{0.2em}\filltoend} \\
Prime & Size & Order& Failures & Qubits & State size & Failures &  Qubits & State size & Failures& Window size \\
\midrule
31 & 5 & 41 & 9.0 &59 & 742 & 8 & 63 & 502 & 4 & 3 \\
61 & 6 & 73 & 5.6 & 68 & 1960 & 3 & 72 & 786 & 2 & 3 \\
97 & 7 & 89 & 4.6 & 77 & 4334 & 2 & 81 & 3442 & 3 & 3 \\
251 & 8 & 271 & 1.8 & 87 & 4476 & 0 & 92 & 5328 & 1 & 4\\
509 & 9 & 503 & 1.0 & 96 & 9036  & 0 & 101 & 14118 & 0 & 4\\
661 & 10 & 665 & 0.8 & 105 & 48682  & 0 & 110 & 48977 & 0 & 4\\
\bottomrule
\end{tabular}}
\end{table*}

\textbf{Discussion.}
The minimum possible qubit count is $8n+\lfloor \lg n\rfloor + 15$, so a full state simulator could run a 3-bit curve (no prime-order 2-bit curve exists) with 40 qubits. To the best of our knowledge, this has not been done.

Our simulator runs larger sizes, allowing us to test the scaling of the error rate with the order of the curve. The probability of each type of failure is upper-bounded by $(2+\lceil\lg(r-1)\rceil)\frac{1}{2(r-1)}$ for an $m$-bit QFT and a curve of order $r$~\cite{roetteler2017quantum}. The types of failures are mutually exclusive. The expected value of failures over 30 trials is given in \cref{tab:ecc-dlog-instances}. Our results suggest it is a slight overestimate, even at these sizes. 

The maximum state size is much more than double the group order. Inspecting the wavefunction during execution shows that much of this superposition consists of points not on the curve with low amplitude, i.e., accumulated point addition errors. Some point addition errors will still result in a 0 state in the auxiliary qubits we measured, and thus contribute to the wavefunction. The superposition size is also much lower than the square of the prime modulus, meaning it was not a superposition over all possible pairs of points. 

Our simulation of windowed point addition represents a logical qubit simulation of the state-of-the-art implementation of Shor's algorithm over elliptic curves. Windowing the point addition is beneficial from a quantum resource perspective, but it actually benefits our simulator as well. We might expect that a $k$-bit window will increase the superposition size by a factor of $2^{k-1}$. Instead, we see little difference. Windowing means fewer point additions, though each one adds more points, and thus should have the same failure rate. The state size at all points is still higher than $2^k$ times the group order.  %

\subsection{Quantum Chemistry}

\textbf{Background.}
The application we consider is finding the ground state energy of a molecular Hamiltonian. Several quantum algorithms are known that address this problem. Here, we focus on two algorithms, both of which apply quantum phase estimation to a subroutine that implements a function of the Hamiltonian $H$.

One such function is the time evolution operator $U_t=e^{-iHt}$, where $t$ denotes the duration of time evolution. Given an implementation of $U_t$ and a trial state $\ket{\tilde\psi_0}$ with a nontrivial overlap with the true ground state $\ket{\psi_0}$, quantum phase estimation outputs an estimate of the ground state energy $E_0$ with probability $p=|\braket{\tilde\psi_0|\psi_0}|^2$.

\textbf{Setup.}
The two algorithms that we use as benchmarks are qubitization~\cite{low2019hamiltonian} and a Trotter-Suzuki based implementation~\cite{lanyon2010towards}. We refer to the papers by \citet{low2019hamiltonian} and \citet{lanyon2010towards}, respectively, for more details. Both algorithms are available in the QDK~\cite{svore2018q}, allowing us to run them using Q\#. We ran them with our sparse simulator and with the full-state simulator from the QDK. As examples, we use an H$_2$ molecule using 2 or 4 spatial orbitals, corresponding to 4 and 8 spin-orbitals, respectively.

\begin{table*}
\caption{Results and parameters of chemistry benchmarks for H$_2$. Tests are parameterized by the number of spin-orbitals (SO), the number of bits in the quantum phase estimation (QPE), and the Trotter step size.}
\label{tab:chem-benchmarks}
{\footnotesize
\begin{tabular}{ccccr rrrr rrr}
\toprule
&&&&&\multicolumn{4}{c}{\filltoend\hspace{0.2em} Sparse Simulator\hspace{0.2em}\filltoend } & \multicolumn{3}{c}{\filltoend\hspace{0.2em} Full state simulator\hspace{0.2em}\filltoend }\\
&&&&& & \multicolumn{3}{c}{Runtime (s)} & \multicolumn{3}{c}{Runtime (s)}\\
Algorithm & SO & QPE Bits & Step size & Qubits & State size & Min. & Median & Max. & Min. & Median & Max.\\
\midrule
Trotter-Suzuki & 4 & 7 & 0.4 & 5 & 4 & 0.20 & 0.20 & 0.46 & 0.22 & 0.23 & 0.44 \\
 		& 8 & 8 & 0.1& 9 & 30 & 9.7 & 9.8 & 9.9 & 9.8 & 9.8 & 10.0 \\
Qubitization & 4 & 7 & n/a& 13 & 64 & 3.6 & 3.6 & 3.8 & 11.5 & 11.5 & 11.7\\
 			& 8 & 8 & n/a & 27 & 24998 & 929 & 958 & 1020 & \multicolumn{3}{c}{>6 days} \\
\bottomrule
\end{tabular}
}
\end{table*}

\textbf{Discussion.}
Our simulator improves upon the full-state simulation, especially for larger number of orbitals and qubitization, which uses more qubits than a Trotter-Suzuki based implementation. This improvement in runtime is possible because the large qubitization benchmark occupied at most 0.02\% of the state vector.

For the smallest Trotter-Suzuki simulation, only about 1 ms was spent in the simulator; the rest of the time was Q\# overhead. The overhead decreased to 11\% on average for the 8 spin-orbital qubitization benchmarks.

\section{Parallelism}\label{sec:parallelism}

It is nontrivial to add efficient multi-threading support to our simulator. The reasons are that (1) most operations need very little computation for each state in superposition and (2) while the hash map is thread-safe for reading, it is not for writing. As a result, it is to be expected that a multi-threaded implementation provides a minor benefit over the serial implementation.

The first issue can be addressed when there is a long queue of phase/permutation gates. The time to apply all queued permutations is then large relative to a write access to the hash map. We find that processing the queue in parallel on each state is beneficial for long queues and large superpositions. Specifically, we only parallelize if the queue is longer than 64 gates and if there are more than 4096 states in the superposition.

Unfortunately, (1) remains an obstacle for other gates. Consequently, only a fraction of the total runtime is subject to parallelism. For example, we find that the simulator spends approximately 60\% of the total serial runtime on large permutations when simulating Shor's algorithm for factoring and elliptic curve discrete logarithms. From Amdahl's law~\cite{amdahl1967validity}, we thus expect a speedup of at most $2.5\times$.

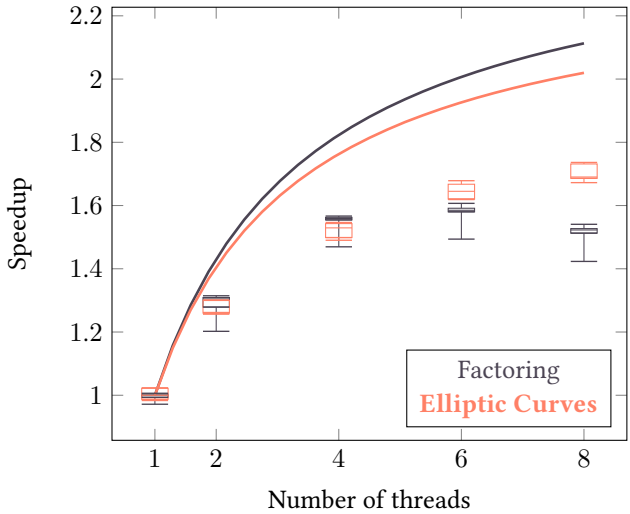
\begin{figure}
\centering
\begin{tikzpicture}
  \begin{axis} [enlarge x limits = 0.1, xtick=data,
  		xlabel = {Number of threads}, ylabel = {Speedup}, legend pos = south east, empty legend]
    \addplot [draw=dataOne, box plot median] table {factorSpeedup.dat};
    \addplot [draw=dataOne, box plot box] table {factorSpeedup.dat};
    \addplot [draw=dataOne, box plot top whisker] table {factorSpeedup.dat};
    \addplot [draw=dataOne, box plot bottom whisker] table {factorSpeedup.dat};
    \addlegendentry[dataOne]{Factoring}
    \addplot [draw=dataTwo, box plot median] table {eccSpeedup.dat};
    \addplot [draw=dataTwo,box plot box] table {eccSpeedup.dat};
    \addplot [draw=dataTwo,box plot top whisker] table {eccSpeedup.dat};
    \addplot [draw=dataTwo,box plot bottom whisker] table {eccSpeedup.dat};
    \addlegendentry[dataTwo]{\textbf{Elliptic Curves}}
    \addplot[draw=dataOne,line width=1pt][domain=1:8] {1/(1-0.602+0.602/x)};
    \addplot[draw=dataTwo,line width=1pt][domain=1:8] {1/(1-0.577+0.577/x)};
   \end{axis}
\end{tikzpicture}%
\caption{Parallel speed-up of Shor's algorithm for elliptic curve discrete logarithm of a 10-bit curve and factoring 961207, relative to the median runtime with a single core. Theoretical speed-up (solid lines) from Amdahl's law is based on the average proportion of time spent on parallelizable gates (0.602 and 0.577, respectively).}\label{fig:parallelism}
\end{figure}

We tested our multi-threaded implementation on the largest parameter sizes for factoring and elliptic curve discrete logarithms, with a number of cores increasing from 1 to 8. We ran these benchmarks on the same machine as in \cref{sec:benchmarks}. We plot the runtime in \cref{fig:parallelism} and compare to the theoretical performance from Amdahl's law~\cite{amdahl1967validity}.

While multi-threading noticeably reduces simulation runtime, our implementation fails to reach the theoretical maximum speedup. We expect that contention to write to the new hash map is causing the issue. Despite spending slightly less time on parallelizable computations, the elliptic curve arithmetic benchmark benefits more from parallelism. This may be another consequence of contention: The factoring circuits used measurement-based uncomputation for comparisons and table look-ups, and measurements force execution of gate queues. In contrast, the elliptic curve arithmetic has no such quantum optimizations built in. Therefore, it can build larger queues of gates, leading to more time spent on processing the gates, rather than competing for write-access to the new hash map.

\section{Conclusion and Outlook}
We introduce a new approach for testing and debugging quantum algorithms. In contrast to the state of the art, our simulator enables complete test runs of optimized implementations of arithmetic with table look-ups and measurement-based uncomputation, and quantum chemistry with qubitization. 

We envision that this simulator can help reduce resource requirements further, as it enables testing of optimizations that could not be tested thus far, and testing many smaller examples even faster. This will also improve the confidence in these optimizations by better testing their accuracy.

Future work on this kind of simulation could incorporate other simulation optimizations, such as separate data structures while states are separable. A hash map which is thread-safe for writing could also improve parallelism, which may be necessary for extremely large simulations that would require a distributed computation.

\begin{acks}
We would like to thank Mathias Soeken for the Q\# implementations of modular arithmetic we used and for help with Q\# integration, Michael Naehrig for helpful references, and Boris Koepf for making this research possible within the constraints of a pandemic.
\end{acks}

\bibliographystyle{ACM-Reference-Format}
\bibliography{bibfile}


\begin{thebibliography}{44}


\ifx \showCODEN    \undefined \def \showCODEN     #1{\unskip}     \fi
\ifx \showDOI      \undefined \def \showDOI       #1{#1}\fi
\ifx \showISBNx    \undefined \def \showISBNx     #1{\unskip}     \fi
\ifx \showISBNxiii \undefined \def \showISBNxiii  #1{\unskip}     \fi
\ifx \showISSN     \undefined \def \showISSN      #1{\unskip}     \fi
\ifx \showLCCN     \undefined \def \showLCCN      #1{\unskip}     \fi
\ifx \shownote     \undefined \def \shownote      #1{#1}          \fi
\ifx \showarticletitle \undefined \def \showarticletitle #1{#1}   \fi
\ifx \showURL      \undefined \def \showURL       {\relax}        \fi
\providecommand\bibfield[2]{#2}
\providecommand\bibinfo[2]{#2}
\providecommand\natexlab[1]{#1}
\providecommand\showeprint[2][]{arXiv:#2}

\bibitem[\protect\citeauthoryear{Aleksandrowicz, Alexander, Barkoutsos, Bello,
  Ben-Haim, Bucher, Cabrera-Hern{\'a}ndez, Carballo-Franquis, Chen, Chen,
  et~al\mbox{.}}{Aleksandrowicz et~al\mbox{.}}{2019}]%
        {aleksandrowicz2019qiskit}
\bibfield{author}{\bibinfo{person}{Gadi Aleksandrowicz},
  \bibinfo{person}{Thomas Alexander}, \bibinfo{person}{Panagiotis Barkoutsos},
  \bibinfo{person}{Luciano Bello}, \bibinfo{person}{Yael Ben-Haim},
  \bibinfo{person}{D Bucher}, \bibinfo{person}{FJ Cabrera-Hern{\'a}ndez},
  \bibinfo{person}{J Carballo-Franquis}, \bibinfo{person}{A Chen},
  \bibinfo{person}{CF Chen}, {et~al\mbox{.}}} \bibinfo{year}{2019}\natexlab{}.
\newblock \showarticletitle{Qiskit: An open-source framework for quantum
  computing}.
\newblock \bibinfo{journal}{\emph{Accessed on: Mar}}  \bibinfo{volume}{16}
  (\bibinfo{year}{2019}).
\newblock


\bibitem[\protect\citeauthoryear{Amdahl}{Amdahl}{1967}]%
        {amdahl1967validity}
\bibfield{author}{\bibinfo{person}{G.M. Amdahl}.}
  \bibinfo{year}{1967}\natexlab{}.
\newblock \showarticletitle{Validity of the single-processor approach to
  achieving large scale computing capabilities}. In
  \bibinfo{booktitle}{\emph{Proceedings of AFIPS Conference}},
  Vol.~\bibinfo{volume}{30}. \bibinfo{pages}{483--485}.
\newblock


\bibitem[\protect\citeauthoryear{Beauregard}{Beauregard}{2003}]%
        {beauregard2003circuit}
\bibfield{author}{\bibinfo{person}{Stephane Beauregard}.}
  \bibinfo{year}{2003}\natexlab{}.
\newblock \showarticletitle{Circuit for Shor's Algorithm Using 2n+3 Qubits}.
\newblock \bibinfo{journal}{\emph{Quantum Info. Comput.}} \bibinfo{volume}{3},
  \bibinfo{number}{2} (\bibinfo{date}{March} \bibinfo{year}{2003}),
  \bibinfo{pages}{175–185}.
\newblock
\showISSN{1533-7146}


\bibitem[\protect\citeauthoryear{Benaloh}{Benaloh}{2020}]%
        {benaloh2020electionguard}
\bibfield{author}{\bibinfo{person}{Josh Benaloh}.}
  \bibinfo{year}{2020}\natexlab{}.
\newblock \showarticletitle{Election{G}uard Specification v0.95}.
\newblock  (\bibinfo{year}{2020}).
\newblock
\newblock
\shownote{\url{https://www.electionguard.vote/spec/Overview/}.}


\bibitem[\protect\citeauthoryear{Berry, Gidney, Motta, McClean, and
  Babbush}{Berry et~al\mbox{.}}{2019}]%
        {berry2019qubitization}
\bibfield{author}{\bibinfo{person}{Dominic~W. Berry}, \bibinfo{person}{Craig
  Gidney}, \bibinfo{person}{Mario Motta}, \bibinfo{person}{Jarrod~R. McClean},
  {and} \bibinfo{person}{Ryan Babbush}.} \bibinfo{year}{2019}\natexlab{}.
\newblock \showarticletitle{Qubitization of {A}rbitrary {B}asis {Q}uantum
  {C}hemistry {L}everaging {S}parsity and {L}ow {R}ank {F}actorization}.
\newblock \bibinfo{journal}{\emph{{Quantum}}}  \bibinfo{volume}{3}
  (\bibinfo{date}{Dec.} \bibinfo{year}{2019}), \bibinfo{pages}{208}.
\newblock
\showISSN{2521-327X}
\urldef\tempurl%
\url{https://doi.org/10.22331/q-2019-12-02-208}
\showDOI{\tempurl}


\bibitem[\protect\citeauthoryear{Bichsel, Baader, Gehr, and Vechev}{Bichsel
  et~al\mbox{.}}{2020}]%
        {bichsel2020silq}
\bibfield{author}{\bibinfo{person}{Benjamin Bichsel},
  \bibinfo{person}{Maximilian Baader}, \bibinfo{person}{Timon Gehr}, {and}
  \bibinfo{person}{Martin Vechev}.} \bibinfo{year}{2020}\natexlab{}.
\newblock \showarticletitle{Silq: A high-level quantum language with safe
  uncomputation and intuitive semantics}. In
  \bibinfo{booktitle}{\emph{Proceedings of the 41st ACM SIGPLAN Conference on
  Programming Language Design and Implementation}}. \bibinfo{pages}{286--300}.
\newblock


\bibitem[\protect\citeauthoryear{Boixo, Isakov, Smelyanskiy, and Neven}{Boixo
  et~al\mbox{.}}{2017}]%
        {boixo2017simulation}
\bibfield{author}{\bibinfo{person}{Sergio Boixo}, \bibinfo{person}{Sergei~V
  Isakov}, \bibinfo{person}{Vadim~N Smelyanskiy}, {and}
  \bibinfo{person}{Hartmut Neven}.} \bibinfo{year}{2017}\natexlab{}.
\newblock \showarticletitle{Simulation of low-depth quantum circuits as complex
  undirected graphical models}.
\newblock \bibinfo{journal}{\emph{arXiv preprint arXiv:1712.05384}}
  (\bibinfo{year}{2017}).
\newblock


\bibitem[\protect\citeauthoryear{Bravyi and Gosset}{Bravyi and Gosset}{2016}]%
        {bravyi2016improved}
\bibfield{author}{\bibinfo{person}{Sergey Bravyi} {and} \bibinfo{person}{David
  Gosset}.} \bibinfo{year}{2016}\natexlab{}.
\newblock \showarticletitle{Improved classical simulation of quantum circuits
  dominated by Clifford gates}.
\newblock \bibinfo{journal}{\emph{Physical review letters}}
  \bibinfo{volume}{116}, \bibinfo{number}{25} (\bibinfo{year}{2016}),
  \bibinfo{pages}{250501}.
\newblock


\bibitem[\protect\citeauthoryear{Cuccaro, Draper, Kutin, and Moulton}{Cuccaro
  et~al\mbox{.}}{2004}]%
        {cuccaro2004new}
\bibfield{author}{\bibinfo{person}{Steven~A. Cuccaro},
  \bibinfo{person}{Thomas~G. Draper}, \bibinfo{person}{Samuel~A. Kutin}, {and}
  \bibinfo{person}{David~Petrie Moulton}.} \bibinfo{year}{2004}\natexlab{}.
\newblock \bibinfo{title}{A new quantum ripple-carry addition circuit}.
\newblock
\newblock
\showeprint{arXiv:quant-ph/0410184}


\bibitem[\protect\citeauthoryear{Dang, Hill, and Hollenberg}{Dang
  et~al\mbox{.}}{2019}]%
        {dang2019optimising}
\bibfield{author}{\bibinfo{person}{Aidan Dang}, \bibinfo{person}{Charles~D
  Hill}, {and} \bibinfo{person}{Lloyd~CL Hollenberg}.}
  \bibinfo{year}{2019}\natexlab{}.
\newblock \showarticletitle{Optimising matrix product state simulations of
  Shor's algorithm}.
\newblock \bibinfo{journal}{\emph{Quantum}}  \bibinfo{volume}{3}
  (\bibinfo{year}{2019}), \bibinfo{pages}{116}.
\newblock


\bibitem[\protect\citeauthoryear{Draper}{Draper}{2000}]%
        {draper2000addition}
\bibfield{author}{\bibinfo{person}{Thomas~G Draper}.}
  \bibinfo{year}{2000}\natexlab{}.
\newblock \showarticletitle{Addition on a quantum computer}.
\newblock \bibinfo{journal}{\emph{arXiv preprint quant-ph/0008033}}
  (\bibinfo{year}{2000}).
\newblock


\bibitem[\protect\citeauthoryear{Gidney}{Gidney}{2017}]%
        {gidney2017factoring}
\bibfield{author}{\bibinfo{person}{Craig Gidney}.}
  \bibinfo{year}{2017}\natexlab{}.
\newblock \bibinfo{title}{Factoring with n+2 clean qubits and n-1 dirty
  qubits}.
\newblock
\newblock
\showeprint{arXiv:1706.07884}


\bibitem[\protect\citeauthoryear{Gidney}{Gidney}{2018}]%
        {gidney2018halving}
\bibfield{author}{\bibinfo{person}{Craig Gidney}.}
  \bibinfo{year}{2018}\natexlab{}.
\newblock \showarticletitle{Halving the cost of quantum addition}.
\newblock \bibinfo{journal}{\emph{{Quantum}}}  \bibinfo{volume}{2}
  (\bibinfo{date}{June} \bibinfo{year}{2018}), \bibinfo{pages}{74}.
\newblock
\showISSN{2521-327X}
\urldef\tempurl%
\url{https://doi.org/10.22331/q-2018-06-18-74}
\showDOI{\tempurl}


\bibitem[\protect\citeauthoryear{Gidney}{Gidney}{2019a}]%
        {gidney2019approximate}
\bibfield{author}{\bibinfo{person}{Craig Gidney}.}
  \bibinfo{year}{2019}\natexlab{a}.
\newblock \bibinfo{title}{Approximate encoded permutations and piecewise
  quantum adders}.
\newblock
\newblock
\showeprint{arXiv:1905.08488}


\bibitem[\protect\citeauthoryear{Gidney}{Gidney}{2019b}]%
        {gidney2019windowed}
\bibfield{author}{\bibinfo{person}{Craig Gidney}.}
  \bibinfo{year}{2019}\natexlab{b}.
\newblock \bibinfo{title}{Windowed quantum arithmetic}.
\newblock
\newblock
\showeprint{arXiv:1905.07682}


\bibitem[\protect\citeauthoryear{Gidney and Eker{\aa}}{Gidney and
  Eker{\aa}}{2019}]%
        {gidney2019factor}
\bibfield{author}{\bibinfo{person}{Craig Gidney} {and} \bibinfo{person}{Martin
  Eker{\aa}}.} \bibinfo{year}{2019}\natexlab{}.
\newblock \showarticletitle{How to factor 2048 bit rsa integers in 8 hours
  using 20 million noisy qubits}.
\newblock \bibinfo{journal}{\emph{arXiv preprint arXiv:1905.09749}}
  (\bibinfo{year}{2019}).
\newblock


\bibitem[\protect\citeauthoryear{Gillmor}{Gillmor}{2016}]%
        {gilmor2016negotiated}
\bibfield{author}{\bibinfo{person}{D. Gillmor}.}
  \bibinfo{year}{2016}\natexlab{}.
\newblock \bibinfo{title}{Negotiated Finite Field {Diffie}-{Hellman} Ephemeral
  Parameters for Transport Layer Security ({TLS})}.
\newblock
\newblock
\urldef\tempurl%
\url{https://tools.ietf.org/html/rfc7919#appendix-A}
\showURL{%
\tempurl}


\bibitem[\protect\citeauthoryear{Green, Lumsdaine, Ross, Selinger, and
  Valiron}{Green et~al\mbox{.}}{2013}]%
        {green2013quipper}
\bibfield{author}{\bibinfo{person}{Alexander~S Green},
  \bibinfo{person}{Peter~LeFanu Lumsdaine}, \bibinfo{person}{Neil~J Ross},
  \bibinfo{person}{Peter Selinger}, {and} \bibinfo{person}{Beno{\^\i}t
  Valiron}.} \bibinfo{year}{2013}\natexlab{}.
\newblock \showarticletitle{Quipper: a scalable quantum programming language}.
  In \bibinfo{booktitle}{\emph{Proceedings of the 34th ACM SIGPLAN conference
  on Programming language design and implementation}}.
  \bibinfo{pages}{333--342}.
\newblock


\bibitem[\protect\citeauthoryear{Griffiths and Niu}{Griffiths and Niu}{1996}]%
        {griffiths1996semiclassical}
\bibfield{author}{\bibinfo{person}{Robert~B. Griffiths} {and}
  \bibinfo{person}{Chi-Sheng Niu}.} \bibinfo{year}{1996}\natexlab{}.
\newblock \showarticletitle{Semiclassical Fourier Transform for Quantum
  Computation}.
\newblock \bibinfo{journal}{\emph{Phys. Rev. Lett.}}  \bibinfo{volume}{76}
  (\bibinfo{date}{Apr} \bibinfo{year}{1996}), \bibinfo{pages}{3228--3231}.
\newblock
Issue 17.
\urldef\tempurl%
\url{https://doi.org/10.1103/PhysRevLett.76.3228}
\showDOI{\tempurl}


\bibitem[\protect\citeauthoryear{Grurl, Fu{\ss}, Hillmich, Burgholzer, and
  Wille}{Grurl et~al\mbox{.}}{2020}]%
        {grurl2020arrays}
\bibfield{author}{\bibinfo{person}{Thomas Grurl}, \bibinfo{person}{J{\"u}rgen
  Fu{\ss}}, \bibinfo{person}{Stefan Hillmich}, \bibinfo{person}{Lukas
  Burgholzer}, {and} \bibinfo{person}{Robert Wille}.}
  \bibinfo{year}{2020}\natexlab{}.
\newblock \showarticletitle{Arrays vs. decision diagrams: A case study on
  quantum circuit simulators}. In \bibinfo{booktitle}{\emph{2020 IEEE 50th
  International Symposium on Multiple-Valued Logic (ISMVL)}}. IEEE,
  \bibinfo{pages}{176--181}.
\newblock


\bibitem[\protect\citeauthoryear{H{\"a}ner, Jaques, Naehrig, Roetteler, and
  Soeken}{H{\"a}ner et~al\mbox{.}}{2020}]%
        {haner2020improved}
\bibfield{author}{\bibinfo{person}{Thomas H{\"a}ner}, \bibinfo{person}{Samuel
  Jaques}, \bibinfo{person}{Michael Naehrig}, \bibinfo{person}{Martin
  Roetteler}, {and} \bibinfo{person}{Mathias Soeken}.}
  \bibinfo{year}{2020}\natexlab{}.
\newblock \showarticletitle{Improved quantum circuits for elliptic curve
  discrete logarithms}. In \bibinfo{booktitle}{\emph{International Conference
  on Post-Quantum Cryptography}}. Springer, \bibinfo{pages}{425--444}.
\newblock


\bibitem[\protect\citeauthoryear{H\"{a}ner, Roetteler, and Svore}{H\"{a}ner
  et~al\mbox{.}}{2017}]%
        {haner2017factoring}
\bibfield{author}{\bibinfo{person}{Thomas H\"{a}ner}, \bibinfo{person}{Martin
  Roetteler}, {and} \bibinfo{person}{Krysta~M. Svore}.}
  \bibinfo{year}{2017}\natexlab{}.
\newblock \showarticletitle{Factoring Using 2n + 2 Qubits with Toffoli Based
  Modular Multiplication}.
\newblock \bibinfo{journal}{\emph{Quantum Info. Comput.}} \bibinfo{volume}{17},
  \bibinfo{number}{7–8} (\bibinfo{date}{June} \bibinfo{year}{2017}),
  \bibinfo{pages}{673–684}.
\newblock
\showISSN{1533-7146}


\bibitem[\protect\citeauthoryear{H{\"a}ner and Steiger}{H{\"a}ner and
  Steiger}{2017}]%
        {haner20175}
\bibfield{author}{\bibinfo{person}{Thomas H{\"a}ner} {and}
  \bibinfo{person}{Damian~S Steiger}.} \bibinfo{year}{2017}\natexlab{}.
\newblock \showarticletitle{0.5 petabyte simulation of a 45-qubit quantum
  circuit}. In \bibinfo{booktitle}{\emph{Proceedings of the International
  Conference for High Performance Computing, Networking, Storage and
  Analysis}}. \bibinfo{pages}{1--10}.
\newblock


\bibitem[\protect\citeauthoryear{H{\"a}ner, Steiger, Smelyanskiy, and
  Troyer}{H{\"a}ner et~al\mbox{.}}{2016}]%
        {haner2016high}
\bibfield{author}{\bibinfo{person}{Thomas H{\"a}ner}, \bibinfo{person}{Damian~S
  Steiger}, \bibinfo{person}{Mikhail Smelyanskiy}, {and}
  \bibinfo{person}{Matthias Troyer}.} \bibinfo{year}{2016}\natexlab{}.
\newblock \showarticletitle{High performance emulation of quantum circuits}. In
  \bibinfo{booktitle}{\emph{SC'16: Proceedings of the International Conference
  for High Performance Computing, Networking, Storage and Analysis}}. IEEE,
  \bibinfo{pages}{866--874}.
\newblock


\bibitem[\protect\citeauthoryear{Hasse}{Hasse}{1936}]%
        {hasse1936zur}
\bibfield{author}{\bibinfo{person}{Helmut Hasse}.}
  \bibinfo{year}{1936}\natexlab{}.
\newblock \showarticletitle{Zur Theorie der abstrakten elliptischen
  Funktionenkörper III. Die Struktur des Meromorphismenrings. Die Riemannsche
  Vermutung.}
\newblock  \bibinfo{volume}{1936}, \bibinfo{number}{175}
  (\bibinfo{year}{1936}), \bibinfo{pages}{193--208}.
\newblock
\urldef\tempurl%
\url{https://doi.org/doi:10.1515/crll.1936.175.193}
\showDOI{\tempurl}


\bibitem[\protect\citeauthoryear{Huang, Zhang, Newman, Cai, Gao, Tian, Wu, Xu,
  Yu, Yuan, et~al\mbox{.}}{Huang et~al\mbox{.}}{2020}]%
        {huang2020classical}
\bibfield{author}{\bibinfo{person}{Cupjin Huang}, \bibinfo{person}{Fang Zhang},
  \bibinfo{person}{Michael Newman}, \bibinfo{person}{Junjie Cai},
  \bibinfo{person}{Xun Gao}, \bibinfo{person}{Zhengxiong Tian},
  \bibinfo{person}{Junyin Wu}, \bibinfo{person}{Haihong Xu},
  \bibinfo{person}{Huanjun Yu}, \bibinfo{person}{Bo Yuan}, {et~al\mbox{.}}}
  \bibinfo{year}{2020}\natexlab{}.
\newblock \showarticletitle{Classical simulation of quantum supremacy
  circuits}.
\newblock \bibinfo{journal}{\emph{arXiv preprint arXiv:2005.06787}}
  (\bibinfo{year}{2020}).
\newblock


\bibitem[\protect\citeauthoryear{JavadiAbhari, Patil, Kudrow, Heckey, Lvov,
  Chong, and Martonosi}{JavadiAbhari et~al\mbox{.}}{2015}]%
        {javadiabhari2015scaffcc}
\bibfield{author}{\bibinfo{person}{Ali JavadiAbhari}, \bibinfo{person}{Shruti
  Patil}, \bibinfo{person}{Daniel Kudrow}, \bibinfo{person}{Jeff Heckey},
  \bibinfo{person}{Alexey Lvov}, \bibinfo{person}{Frederic~T Chong}, {and}
  \bibinfo{person}{Margaret Martonosi}.} \bibinfo{year}{2015}\natexlab{}.
\newblock \showarticletitle{ScaffCC: Scalable compilation and analysis of
  quantum programs}.
\newblock \bibinfo{journal}{\emph{Parallel Comput.}}  \bibinfo{volume}{45}
  (\bibinfo{year}{2015}), \bibinfo{pages}{2--17}.
\newblock


\bibitem[\protect\citeauthoryear{Jones, Brown, Bush, and Benjamin}{Jones
  et~al\mbox{.}}{2019}]%
        {jones2019quest}
\bibfield{author}{\bibinfo{person}{Tyson Jones}, \bibinfo{person}{Anna Brown},
  \bibinfo{person}{Ian Bush}, {and} \bibinfo{person}{Simon~C Benjamin}.}
  \bibinfo{year}{2019}\natexlab{}.
\newblock \showarticletitle{QuEST and high performance simulation of quantum
  computers}.
\newblock \bibinfo{journal}{\emph{Scientific reports}} \bibinfo{volume}{9},
  \bibinfo{number}{1} (\bibinfo{year}{2019}), \bibinfo{pages}{1--11}.
\newblock


\bibitem[\protect\citeauthoryear{Lanyon, Whitfield, Gillett, Goggin, Almeida,
  Kassal, Biamonte, Mohseni, Powell, Barbieri, et~al\mbox{.}}{Lanyon
  et~al\mbox{.}}{2010}]%
        {lanyon2010towards}
\bibfield{author}{\bibinfo{person}{Benjamin~P Lanyon}, \bibinfo{person}{James~D
  Whitfield}, \bibinfo{person}{Geoff~G Gillett}, \bibinfo{person}{Michael~E
  Goggin}, \bibinfo{person}{Marcelo~P Almeida}, \bibinfo{person}{Ivan Kassal},
  \bibinfo{person}{Jacob~D Biamonte}, \bibinfo{person}{Masoud Mohseni},
  \bibinfo{person}{Ben~J Powell}, \bibinfo{person}{Marco Barbieri},
  {et~al\mbox{.}}} \bibinfo{year}{2010}\natexlab{}.
\newblock \showarticletitle{Towards quantum chemistry on a quantum computer}.
\newblock \bibinfo{journal}{\emph{Nature chemistry}} \bibinfo{volume}{2},
  \bibinfo{number}{2} (\bibinfo{year}{2010}), \bibinfo{pages}{106--111}.
\newblock


\bibitem[\protect\citeauthoryear{Low, Bauman, Granade, Peng, Wiebe, Bylaska,
  Wecker, Krishnamoorthy, Roetteler, Kowalski, Troyer, and Baker}{Low
  et~al\mbox{.}}{2019}]%
        {low2019qsharp}
\bibfield{author}{\bibinfo{person}{Guang~Hao Low}, \bibinfo{person}{Nicholas~P.
  Bauman}, \bibinfo{person}{Christopher~E. Granade}, \bibinfo{person}{Bo Peng},
  \bibinfo{person}{Nathan Wiebe}, \bibinfo{person}{Eric~J. Bylaska},
  \bibinfo{person}{Dave Wecker}, \bibinfo{person}{Sriram Krishnamoorthy},
  \bibinfo{person}{Martin Roetteler}, \bibinfo{person}{Karol Kowalski},
  \bibinfo{person}{Matthias Troyer}, {and} \bibinfo{person}{Nathan~A. Baker}.}
  \bibinfo{year}{2019}\natexlab{}.
\newblock \bibinfo{title}{Q\# and NWChem: Tools for Scalable Quantum Chemistry
  on Quantum Computers}.
\newblock
\newblock
\showeprint{arXiv:1904.01131}


\bibitem[\protect\citeauthoryear{Low and Chuang}{Low and Chuang}{2019}]%
        {low2019hamiltonian}
\bibfield{author}{\bibinfo{person}{Guang~Hao Low} {and}
  \bibinfo{person}{Isaac~L Chuang}.} \bibinfo{year}{2019}\natexlab{}.
\newblock \showarticletitle{Hamiltonian simulation by qubitization}.
\newblock \bibinfo{journal}{\emph{Quantum}}  \bibinfo{volume}{3}
  (\bibinfo{year}{2019}), \bibinfo{pages}{163}.
\newblock


\bibitem[\protect\citeauthoryear{Markov, Fatima, Isakov, and Boixo}{Markov
  et~al\mbox{.}}{2018}]%
        {markov2018quantum}
\bibfield{author}{\bibinfo{person}{Igor~L Markov}, \bibinfo{person}{Aneeqa
  Fatima}, \bibinfo{person}{Sergei~V Isakov}, {and} \bibinfo{person}{Sergio
  Boixo}.} \bibinfo{year}{2018}\natexlab{}.
\newblock \showarticletitle{Quantum supremacy is both closer and farther than
  it appears}.
\newblock \bibinfo{journal}{\emph{arXiv preprint arXiv:1807.10749}}
  (\bibinfo{year}{2018}).
\newblock


\bibitem[\protect\citeauthoryear{McArdle, Endo, Aspuru-Guzik, Benjamin, and
  Yuan}{McArdle et~al\mbox{.}}{2020}]%
        {mcardle2020quantum}
\bibfield{author}{\bibinfo{person}{Sam McArdle}, \bibinfo{person}{Suguru Endo},
  \bibinfo{person}{Alan Aspuru-Guzik}, \bibinfo{person}{Simon~C Benjamin},
  {and} \bibinfo{person}{Xiao Yuan}.} \bibinfo{year}{2020}\natexlab{}.
\newblock \showarticletitle{Quantum computational chemistry}.
\newblock \bibinfo{journal}{\emph{Reviews of Modern Physics}}
  \bibinfo{volume}{92}, \bibinfo{number}{1} (\bibinfo{year}{2020}),
  \bibinfo{pages}{015003}.
\newblock


\bibitem[\protect\citeauthoryear{Nielsen and Chuang}{Nielsen and
  Chuang}{2002}]%
        {nielsen2002quantum}
\bibfield{author}{\bibinfo{person}{Michael~A Nielsen} {and}
  \bibinfo{person}{Isaac Chuang}.} \bibinfo{year}{2002}\natexlab{}.
\newblock \bibinfo{title}{Quantum computation and quantum information}.
\newblock
\newblock


\bibitem[\protect\citeauthoryear{Roetteler, Naehrig, Svore, and
  Lauter}{Roetteler et~al\mbox{.}}{2017}]%
        {roetteler2017quantum}
\bibfield{author}{\bibinfo{person}{Martin Roetteler}, \bibinfo{person}{Michael
  Naehrig}, \bibinfo{person}{Krysta~M. Svore}, {and} \bibinfo{person}{Kristin
  Lauter}.} \bibinfo{year}{2017}\natexlab{}.
\newblock \showarticletitle{Quantum Resource Estimates for Computing Elliptic
  Curve Discrete Logarithms}. In \bibinfo{booktitle}{\emph{Advances in
  Cryptology -- ASIACRYPT 2017}}, \bibfield{editor}{\bibinfo{person}{Tsuyoshi
  Takagi} {and} \bibinfo{person}{Thomas Peyrin}} (Eds.).
  \bibinfo{publisher}{Springer International Publishing},
  \bibinfo{address}{Cham}, \bibinfo{pages}{241--270}.
\newblock


\bibitem[\protect\citeauthoryear{Shor}{Shor}{1994}]%
        {shor1994algorithms}
\bibfield{author}{\bibinfo{person}{Peter~W Shor}.}
  \bibinfo{year}{1994}\natexlab{}.
\newblock \showarticletitle{Algorithms for quantum computation: discrete
  logarithms and factoring}. In \bibinfo{booktitle}{\emph{Proceedings 35th
  annual symposium on foundations of computer science}}. Ieee,
  \bibinfo{pages}{124--134}.
\newblock


\bibitem[\protect\citeauthoryear{Skarupke}{Skarupke}{2018}]%
        {skarupke2018hash}
\bibfield{author}{\bibinfo{person}{Malte Skarupke}.}
  \bibinfo{year}{2018}\natexlab{}.
\newblock \bibinfo{title}{Flat hash map}.
\newblock
  \bibinfo{howpublished}{\url{https://github.com/skarupke/flat_hash_map}}.
\newblock


\bibitem[\protect\citeauthoryear{Smelyanskiy, Sawaya, and
  Aspuru-Guzik}{Smelyanskiy et~al\mbox{.}}{2016}]%
        {smelyanskiy2016qhipster}
\bibfield{author}{\bibinfo{person}{Mikhail Smelyanskiy},
  \bibinfo{person}{Nicolas~PD Sawaya}, {and} \bibinfo{person}{Al{\'a}n
  Aspuru-Guzik}.} \bibinfo{year}{2016}\natexlab{}.
\newblock \showarticletitle{qHiPSTER: The quantum high performance software
  testing environment}.
\newblock \bibinfo{journal}{\emph{arXiv preprint arXiv:1601.07195}}
  (\bibinfo{year}{2016}).
\newblock


\bibitem[\protect\citeauthoryear{Steiger, H{\"a}ner, and Troyer}{Steiger
  et~al\mbox{.}}{2018}]%
        {steiger2018projectq}
\bibfield{author}{\bibinfo{person}{Damian~S Steiger}, \bibinfo{person}{Thomas
  H{\"a}ner}, {and} \bibinfo{person}{Matthias Troyer}.}
  \bibinfo{year}{2018}\natexlab{}.
\newblock \showarticletitle{ProjectQ: an open source software framework for
  quantum computing}.
\newblock \bibinfo{journal}{\emph{Quantum}}  \bibinfo{volume}{2}
  (\bibinfo{year}{2018}), \bibinfo{pages}{49}.
\newblock


\bibitem[\protect\citeauthoryear{Svore, Geller, Troyer, Azariah, Granade, Heim,
  Kliuchnikov, Mykhailova, Paz, and Roetteler}{Svore et~al\mbox{.}}{2018}]%
        {svore2018q}
\bibfield{author}{\bibinfo{person}{Krysta Svore}, \bibinfo{person}{Alan
  Geller}, \bibinfo{person}{Matthias Troyer}, \bibinfo{person}{John Azariah},
  \bibinfo{person}{Christopher Granade}, \bibinfo{person}{Bettina Heim},
  \bibinfo{person}{Vadym Kliuchnikov}, \bibinfo{person}{Mariia Mykhailova},
  \bibinfo{person}{Andres Paz}, {and} \bibinfo{person}{Martin Roetteler}.}
  \bibinfo{year}{2018}\natexlab{}.
\newblock \showarticletitle{Q\# enabling scalable quantum computing and
  development with a high-level dsl}. In \bibinfo{booktitle}{\emph{Proceedings
  of the Real World Domain Specific Languages Workshop 2018}}.
  \bibinfo{pages}{1--10}.
\newblock


\bibitem[\protect\citeauthoryear{{Van Meter} and Itoh}{{Van Meter} and
  Itoh}{2005}]%
        {vanmeter2005fast}
\bibfield{author}{\bibinfo{person}{Rodney {Van Meter}} {and}
  \bibinfo{person}{{Kohei M.} Itoh}.} \bibinfo{year}{2005}\natexlab{}.
\newblock \showarticletitle{Fast quantum modular exponentiation}.
\newblock \bibinfo{journal}{\emph{Physical Review A}} \bibinfo{volume}{71},
  \bibinfo{number}{5} (\bibinfo{date}{May} \bibinfo{year}{2005}).
\newblock
\showISSN{2469-9926}
\urldef\tempurl%
\url{https://doi.org/10.1103/PhysRevA.71.052320}
\showDOI{\tempurl}
\newblock
\shownote{Copyright: Copyright 2012 Elsevier B.V., All rights reserved.}


\bibitem[\protect\citeauthoryear{Villalonga, Lyakh, Boixo, Neven, Humble,
  Biswas, Rieffel, Ho, and Mandr{\`a}}{Villalonga et~al\mbox{.}}{2020}]%
        {villalonga2020establishing}
\bibfield{author}{\bibinfo{person}{Benjamin Villalonga},
  \bibinfo{person}{Dmitry Lyakh}, \bibinfo{person}{Sergio Boixo},
  \bibinfo{person}{Hartmut Neven}, \bibinfo{person}{Travis~S Humble},
  \bibinfo{person}{Rupak Biswas}, \bibinfo{person}{Eleanor~G Rieffel},
  \bibinfo{person}{Alan Ho}, {and} \bibinfo{person}{Salvatore Mandr{\`a}}.}
  \bibinfo{year}{2020}\natexlab{}.
\newblock \showarticletitle{Establishing the quantum supremacy frontier with a
  281 pflop/s simulation}.
\newblock \bibinfo{journal}{\emph{Quantum Science and Technology}}
  \bibinfo{volume}{5}, \bibinfo{number}{3} (\bibinfo{year}{2020}),
  \bibinfo{pages}{034003}.
\newblock


\bibitem[\protect\citeauthoryear{Zalka}{Zalka}{2006}]%
        {zalka2006shor}
\bibfield{author}{\bibinfo{person}{Christof Zalka}.}
  \bibinfo{year}{2006}\natexlab{}.
\newblock \showarticletitle{Shor's algorithm with fewer (pure) qubits}.
\newblock \bibinfo{journal}{\emph{arXiv preprint quant-ph/0601097}}
  (\bibinfo{year}{2006}).
\newblock


\bibitem[\protect\citeauthoryear{Zulehner and Wille}{Zulehner and
  Wille}{2018}]%
        {zulehner2018advanced}
\bibfield{author}{\bibinfo{person}{Alwin Zulehner} {and}
  \bibinfo{person}{Robert Wille}.} \bibinfo{year}{2018}\natexlab{}.
\newblock \showarticletitle{Advanced simulation of quantum computations}.
\newblock \bibinfo{journal}{\emph{IEEE Transactions on Computer-Aided Design of
  Integrated Circuits and Systems}} \bibinfo{volume}{38}, \bibinfo{number}{5}
  (\bibinfo{year}{2018}), \bibinfo{pages}{848--859}.
\newblock


\end{thebibliography}

\end{document}